\documentclass{article}
\usepackage{spconf,amsmath,graphicx,hyperref}

\usepackage{booktabs,multirow,pifont}
\usepackage{multirow}
\usepackage{multicol}
\usepackage{bm}
\usepackage{amssymb}
\usepackage{booktabs}
\usepackage{enumitem}
\usepackage{amsthm,amsmath,amssymb}
\usepackage{tabularx}
\usepackage{mathrsfs}
\usepackage{booktabs}
\usepackage{pifont}
\usepackage[table]{xcolor}
\definecolor{ForestGreen}{RGB}{34,139,34}
\newcommand{\cmark}{\textcolor{ForestGreen}{\ding{51}}} 
\newcommand{\xmark}{\textcolor{red}{\ding{55}}}         
\usepackage{array} 
\usepackage{hyperref}
\usepackage{array,makecell,booktabs,multirow}
\newcolumntype{L}[1]{>{\raggedright\arraybackslash}p{#1}}
\usepackage{multirow,booktabs,array}

\usepackage[table]{xcolor}

\title{MULTI-Bench: A Multi-Turn Interactive Benchmark for Assessing Emotional Intelligence ability of Spoken Dialogue Models}

\name{
    \begin{tabular}[t]{c} 
         Yayue Deng$^{1*}$,
         Guoqiang Hu$^{1*,2}$,
         Haiyang Sun$^1$,
         Xiangyu Zhang$^{1,3}$,
         Haoyang Zhang$^{1,4}$, \\ 
         \textit{Fei Tian$^1$,
         Xuerui Yang$^1$,
         Gang Yu$^{1}$,
         Eng Siong Chng$^{2,\ddagger}$}
    \end{tabular} 
    \thanks{* Both authors contributed equally to this research.}
    \thanks{$\dagger$ Work done during internship at StepFun.}
    \thanks{$\ddagger$ Corresponding author.}
}

\address{
  $^1$StepFun Inc, Shanghai, China\\
  $^2$Nanyang Technological University, Singapore
  $^3$ The University of New South Wales, Sydney, Australia\\ 
  $^4$Peking University,Beijing, China
}
\begin{document}
\ninept
\maketitle
\begin{abstract}
Spoken Dialogue Models (SDMs) have advanced rapidly, yet their ability to sustain genuinely interactive multi-turn conversations remains underexplored, as most benchmarks focus on single-turn exchanges. We introduce Multi-Bench, the first benchmark explicitly designed to evaluate SDMs in multi-turn interactive dialogue with an emphasis on emotional intelligence. Multi-Bench employs a hierarchical structure with a basic track for emotion understanding and reasoning and an advanced track for emotion support and application. It comprises five carefully designed tasks and about 3.2K samples, ranging from emotion recognition to complex reasoning and interactive dialogue, supported by a reproducible evaluation framework. We evaluate six representative SDMs on eight subsets of Multi-Bench. Results show that while current SDMs achieve good performance on basic understanding tasks, they still have room for improvement in advanced multi-turn interactive dialogue and reasoning-related tasks, particularly in emotion awareness and application. 

\end{abstract}
\section{Introduction}
\label{sec:intro}
Spoken dialogue models (SDMs), which process speech and generate intelligent audio responses and are exemplified by GPT-4o~\cite{gpt4o}, have recently become a central focus in auditory AI research. More recently, several SDMs~\cite{xu2025qwen25omnitechnicalreport,ding2025kimi,huang2025stepaudioaqaa} have demonstrated performance approaching that of GPT-4o. As these models advance, expectations have shifted far beyond simple speech recognition to include higher-level tasks such as audio-grounded reasoning and interactive dialogue, which in turn pose new challenges for evaluation. Hence, several studies~\cite{yan2025uro,ma2025c3,cheng2025voxdialogue} have assessed models not only on basic understanding tasks, such as Automatic Speech Recognition, commonsense knowledge, or mathematical questions, but also on their performance in the chat dimension, addressing complex real-world scenarios. For instance, URO-Bench~\cite{yan2025uro} evaluates understanding, reasoning, and oral interaction through two tracks: a basic track for simple daily conversations and a pro track for advanced tasks such as emotion recognition, multilingual processing, and multi-turn dialogues. Similarly, AIR-Bench~\cite{yang2024air} provides foundation and chat benchmarks to assess models on diverse audio comprehension and instruction following; it employs free-form outputs and an LLM-based judge to score curated open-ended audio questions in the chat dimension. In contrast, SpokenWOZ~\cite{si2023spokenwoz} focuses on task-oriented dialogue systems for practical goals such as flight booking and restaurant reservation. Meanwhile, SD-Eval~\cite{ao2024sd} emphasizes paralinguistic and environmental information across four aspects: emotion, accent, age, and background sounds. Furthermore, C$^3$Benchmark~\cite{ma2025c3} investigates dialogue understanding in terms of ambiguity and context dependency, with English and Chinese tasks covering phonological and semantic ambiguity as well as context-dependent phenomena such as omission, coreference, and multi-turn interaction. ContextDialog~\cite{contextdialogue} measures recall through spoken QA pairs derived from existing dialogues, requiring models to reference previously mentioned information.

\begin{table}[t]
\centering
\small
\resizebox{0.97\columnwidth}{!}{
\begin{tabular}{lccc cc}
\toprule
\textbf{Benchmark} & \textbf{Multi-Turn} & \textbf{Interactive} & \multicolumn{2}{c}{\textbf{Assessed Modalities}} \\
\cmidrule(lr){4-5}
 &  &  & \textbf{Text} & \textbf{Speech} \\
\midrule
VoiceBench~\cite{chen2024voicebench}        & \xmark & \xmark & \cmark & \xmark \\
AIR-Bench~\cite{yang2024air}         & \xmark & \xmark & \cmark & \xmark \\
ADU-Bench~\cite{adubmk}         & \xmark & \xmark & \cmark & \xmark \\
SD-Eval~\cite{ao2024sd}           & \xmark & \xmark & \cmark & \xmark \\
SOVA-Bench~\cite{hou2025sova}        & \xmark & \xmark & \cmark & \xmark \\
ContextDialog~\cite{contextdialogue}     & \cmark & \xmark & \cmark & \xmark \\
C$^3$Benchmark~\cite{ma2025c3}    & \cmark & \xmark & \cmark & \xmark \\
SpokenWOZ~\cite{si2023spokenwoz}         & \cmark & \xmark & \cmark & \xmark \\
URO-Bench~\cite{yan2025uro}         & \cmark & \xmark & \cmark & \cmark \\
VoxDialogue~\cite{cheng2025voxdialogue}       & \cmark & \xmark & \cmark & \cmark \\
Multi-Bench (ours)  & \cmark & \cmark & \cmark & \cmark \\
\bottomrule
\end{tabular}
}
\caption{Comparison with existing SDM benchmarks. The \textit{Assessed Modalities} columns show whether the benchmark evaluates dialogue in text or speech. For speech, subjective human ratings such as MOS are excluded.}
\label{tab:slm-benchmarks}
\vspace{-0.55cm}
\end{table}

\begin{figure*}[t]
  \centering
  \includegraphics[width=\linewidth]{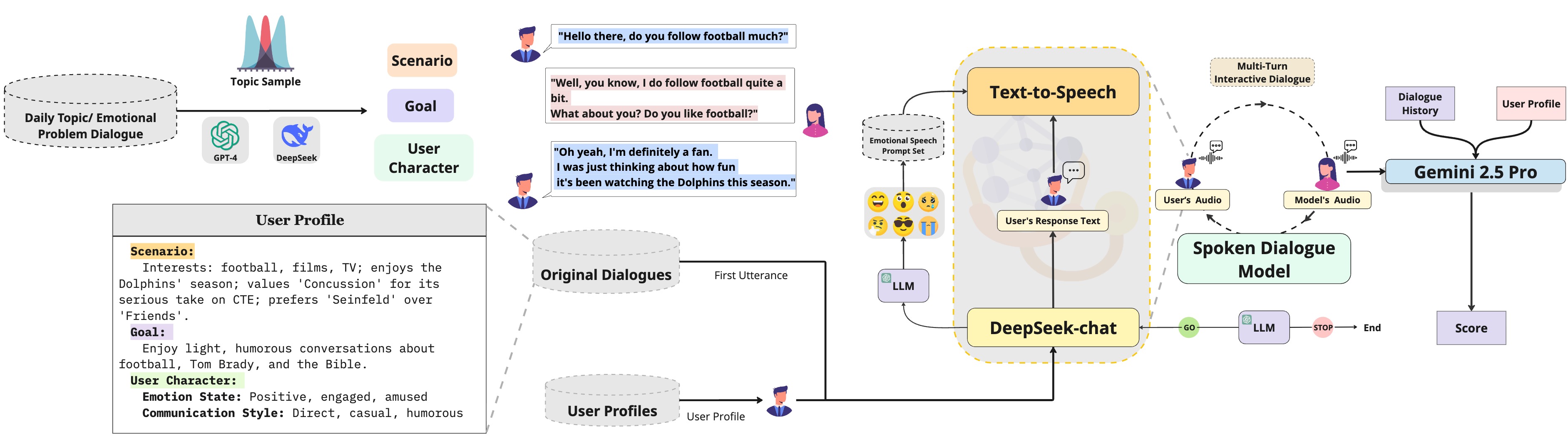}
  \caption{Illustration of the proposed multi-turn interactive evaluation framework in Multi-Bench.}
  \label{pic:overview}
  \vspace{-0.4cm}
\end{figure*}

However, most existing evaluation efforts for SDMs have primarily concentrated on single-turn interactions, while their ability to sustain multi-turn conversations has often been overlooked.
As shown in Table~\ref{tab:slm-benchmarks}, current benchmarks exhibit significant limitations in evaluating multi-turn conversational capabilities. First, most benchmarks~\cite{adubmk,chen2024voicebench,yang2024air,ao2024sd} consist solely of audio queries designed to assess general capabilities rather than contextual dialogue. Second, the majority of these benchmarks~\cite{contextdialogue,yang2024air,ao2024sd} evaluate performance based solely on textual metrics, such as calculating the accuracy of text responses, while ignoring other crucial modalities like audio context and prosodic features. Besides, some multi-turn benchmarks~\cite{ma2025c3,contextdialogue} rely exclusively on narrow metrics such as recall. For example, the multi-turn subtasks in C$^3$-Bench~\cite{ma2025c3} and ContextDialog~\cite{contextdialogue} merely test whether a model can repeat an initial question after several turns, resulting in a limited evaluation scope. Although benchmarks like URO-Bench~\cite{yan2025uro} attempt to incorporate multi-turn dialogues, the interactions are not truly conversational. Instead, they are often concatenations of independent single-turn exchanges. Finally, none of the existing benchmarks evaluate SDMs in interactive multi-turn dialogue.
To address this, we introduce Multi-Bench, the first benchmark designed to evaluate the emotional intelligence of SDMs through multi-turn interactive dialogues. It features a two-tier evaluation structure: a basic track for emotion understanding and reasoning, and an advanced track for emotion support and application. The benchmark includes five carefully designed tasks and a reproducible evaluation framework, filling a critical gap in multi-turn conversational assessment. Our main contributions are as follows:
\begin{itemize}[topsep=1pt, partopsep=0pt, itemsep=0pt, parsep=0pt, leftmargin=*]
    \item We propose \textbf{Multi-Bench}, the first benchmark for evaluating SDMs in genuine multi-turn dialogues, addressing the lack of interactive evaluation in existing single-turn or pseudo multi-turn benchmarks. 
    \item We design a \textbf{hierarchical evaluation framework} with basic and advanced tracks, along with five tailored tasks, to enable fine-grained and comprehensive assessment of emotional intelligence. 
    \item We perform extensive experiments with leading SDMs to validate Multi-Bench, providing empirical insights into their capabilities and limitations in sustaining emotionally intelligent multi-turn dialogue. 
\end{itemize}

\begin{table}[t]
\centering
\caption{Statistics for Multi-Bench. LM Judge and ALLM Judge denote the use of languag model and audio-aware large language model to assess responses.}
\label{tab:statistics multi-bench}
\resizebox{\columnwidth}{!}{%
\begin{tabular}{
    >{\centering\arraybackslash}p{3.2cm}  
    >{\raggedright\arraybackslash}p{3.5cm}  
    >{\raggedright\arraybackslash}p{1.7cm}  
    >{\raggedright\arraybackslash}p{2.0cm}  
    r                                      
    c c                                    
    >{\centering\arraybackslash}p{1.2cm} 
}
\toprule
\multirow{2}{*}{\textbf{Dimension}} &
\multirow{2}{*}{\textbf{Task}} &
\multirow{2}{*}{\textbf{Source Data}} &
\multirow{2}{*}{\textbf{Format}} &
\multirow{2}{*}{\textbf{Num}} &
\multicolumn{2}{c}{\textbf{Eval Level}} &
\multirow{2}{*}{\textbf{Metrics}} \\
\cmidrule(lr){6-7}
& & & & & \textbf{Utt} & \textbf{Dia} & \\
\midrule
\midrule

\multirow{2}{*}{\parbox{3.2cm}{\centering Emotion Understanding \\ and Recognition}}
& Emotion Recognition & UnderEmotion & Semi-Open  & 216 & \cmark & \xmark & \multirow{1}{*}{LM Judge} \\

& Paralinguistic Recognition & NVSpeech & Multi-Choice  & 800 & \cmark & \xmark  &  ACC  \\ 
\cmidrule(lr){2-8}
\multirow{5}{*}{\parbox{3cm}{\centering Emotion Reasoning \\ and Application}}
& Style Inference & StyleTalk & Single-Choice  & 586 & \cmark & \xmark &  \multirow{2}{*}{\parbox{1cm}{\centering ACC}}\\
& Emotion Inference & NVSpeech & Semi-Open & 360 & \cmark & \xmark &  \\
\cmidrule(lr){2-8}

& \multirow{4}{*}{Interactive Dialogue} & PsyQA & Open-domain & 250 & \cmark & \cmark & \multirow{4}{*}{\parbox{1cm}{\centering ALLM \\ Judge}} \\
&  & PsyDTCorpus & Open-domain & 250 & \cmark & \cmark &  \\
&  & MultiDialog & Open-domain & 500 & \cmark & \cmark &  \\
&  & NVSpeech & Open-domain & 500 & \cmark & \cmark &   \\
\bottomrule
\end{tabular}%
}
\vspace{-0.45cm}
\end{table}

\vspace{-0.4cm}
\section{Multi-Bench}
\label{sec:Multi-bench}

Multi-Bench is distinguished from existing benchmarks by three key characteristics: (1) Multi-Bench is the first benchmark to evaluate SDMs' emotional intelligence in an interactive multi-turn conversation scenario. (2) Multi-Bench evaluates emotional intelligence systematically using a hierarchical taxonomy: emotion understanding and reasoning, emotion support and application. (3) Multi-Bench comprehensively evaluates the utterances generated by SDMs from both linguistic and acoustic perspectives, at both the utterance level and the conversation level.

\vspace{-0.4cm}
\subsection{Overview}
\label{subsec:overview}
Emotional Intelligence (EI), introduced by Salovey and Mayer~\cite{Emotional_Intelligence}, describes the ability to perceive, interpret, and regulate one's own and others' emotions, and to use this understanding to guide reasoning and behavior. Subsequent work by Schuller et al.~\cite{schuller2018age} expanded this concept to include emotional adaptation and problem-solving. More recently, Sabour et al.~\cite{sabour2024emobench} proposed EMOBENCH, a benchmark that frames EI in machines through two core dimensions: Emotional Understanding and Emotional Application. While definitions vary, a common consensus remains: EI involves not only accurate emotion perception and tracking, but also the effective application of emotional knowledge to support reasoning, regulation, and decision-making. Building on these foundations, Multi-Bench operationalizes the evaluation of EI in SDMs through a structured multi-turn dialogue framework. As shown in Table~\ref{tab:statistics multi-bench}, we measure the EI of SDMs from two core dimensions:
\begin{itemize}[topsep=2pt, partopsep=0pt, itemsep=0pt, parsep=0pt, leftmargin=*]
\item \textbf{Emotion Understanding and Recognition: }This perspective emphasizes the model’s ability to detect and categorize emotions and paralinguistic cues at the utterance level. To this end, we include tasks such as Emotion Recognition and Paralinguistic Recognition. To increase the diversity of evaluation, these tasks are presented in both single-choice and multi-choice formats.
\item \textbf{Emotion Reasoning and Application:} Beyond recognition, this dimension emphasizes how models interpret nuanced emotional states and respond appropriately in multi-turn interactive dialogue. It encompasses three tasks: Style Inference and Emotion Inference, which require models to capture subtle affective meanings, and the Interactive Dialogue task, which evaluates the ability to sustain emotionally intelligent multi-turn conversations. 
\end{itemize}
\begin{figure*}[t]
  \centering
  \includegraphics[width=0.8\linewidth]{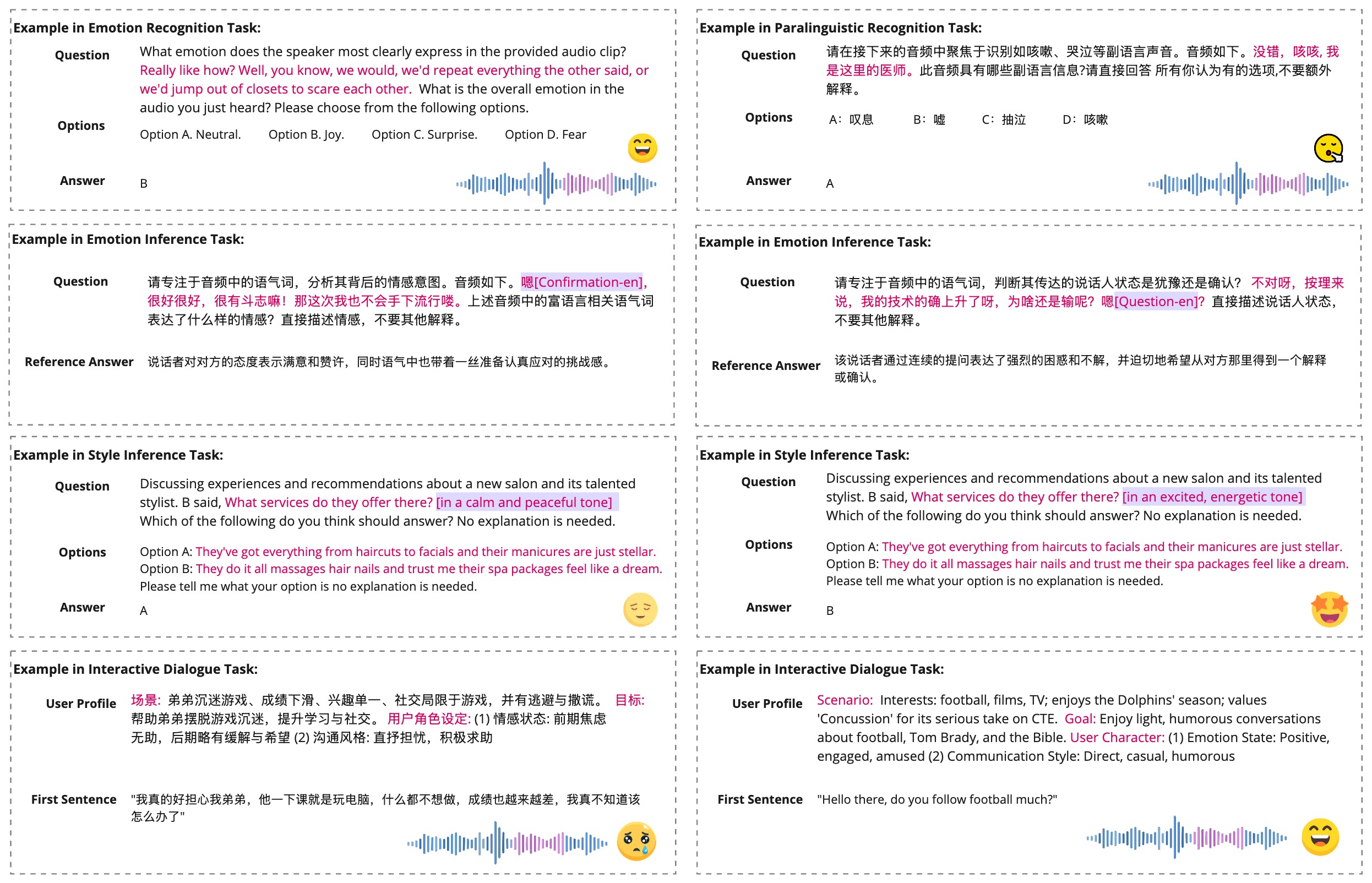}
  \caption{Example data for the five sub-tasks in Multi-Bench: Emotion Recognition, Paralinguistic Recognition, Emotion Inference, Style Inference, and Interactive Dialogue.}
  \label{pic:data_example}
  \vspace{-0.6cm}
\end{figure*}
Overall, Multi-Bench comprises 3,212 samples covering tasks from basic emotion recognition to complex reasoning and interactive dialogue, drawing on datasets such as UnderEmotion, NVSpeech, PsyQA, PsyDTCorpus, and MultiDialog to span diverse topics from everyday conversation to psychological support in both single-turn and multi-turn settings. Our code and data will be open-sourced\footnote{https://mia11939.github.io/MULTI-BENCH/demo.html}.
\begin{table}[t]
\centering
\small
\caption{Performance comparison of models on emotion and paralinguistic tasks. Emotion denotes emotion recognition, Paralinguistic denotes paralinguistic recognition task; Paralinguistic$_{easy}$ accepts partial correctness, while Paralinguistic$_{hard}$ requires full correctness.}
\resizebox{\linewidth}{!}{
\begin{tabular}{lccc|cc}
\toprule
\textbf{Models} & \multicolumn{3}{c|}{\textbf{Understanding and Recognition}} & \multicolumn{2}{c}{\textbf{Reasoning and Application}} \\
\cmidrule(lr){2-4} \cmidrule(lr){5-6}
 & Emotion & Paralinguistic$_{easy}$ & Paralinguistic$_{hard}$ & Style Inference & Emotion Inference \\
\midrule
GLM 4 Voice        & 62.38\% & 46.24\% & 10.46\% & 42.15\% & 36.67\% \\
Qwen 2.5 Omni      & 58.76\% & \underline{68.56}\% & 8.59\%  & 45.56\% & 33.89\% \\
Kimi Audio         & 63.86\% & 66.35\% & 6.98\%  & 55.29\% & 35.28\% \\
Step-Audio-AQAA    & 56.76\% & 56.00\% & 5.69\%  & 51.37\% & 26.94\% \\
Step Audio 2       & \textbf{70.80}\% & 65.43\% & \underline{13.20}\% & \underline{56.14}\% & \underline{40.00}\% \\
GPT-4o              & \underline{65.65}\% & \textbf{70.97\%} & \textbf{17.84\%} & \textbf{64.16\%} & \textbf{42.78\%} \\
\bottomrule
\end{tabular}
}
\label{tab:understanding}
\vspace{-0.4cm}
\end{table}


\begin{table*}[t]
\centering
\caption{Performance comparison of models on \textbf{interactive dialogue tasks} over PsyQA, PsyDTCorpus, Multidialogue, and NVSpeech. Psy denotes the combined PsyQA and PsyDTCorpus datasets. Gemini refers to the utterance-level scores obtained using Gemini 2.5 Pro (1–5 scale), while Global represents the dialogue-level scores produced by DeepSeek.}
\small
\resizebox{0.95\textwidth}{!}{
\begin{tabular}{lccc|ccc|ccc|cccc}
\toprule
\textbf{Models} 
& \multicolumn{3}{c|}{\textbf{Psy}} 
& \multicolumn{3}{c|}{\textbf{Multidialogue}} 
& \multicolumn{3}{c|}{\textbf{NVSpeech}} 
& \multicolumn{3}{c}{\textbf{Avg}} \\
\cmidrule(lr){2-4} \cmidrule(lr){5-7} \cmidrule(lr){8-10} \cmidrule(lr){11-13}
& Gemini & Global & UTMOS & Gemini & Global & UTMOS & Gemini & Global & UTMOS & Gemini & Global & UTMOS \\
\midrule
Qwen 2.5 Omni    & 3.457 & 3.66  & 3.19 & 3.054 & 3.94 & 4.43 & 3.345 & 3.66  & 3.21 & 3.285 & 3.75 & 3.61  \\
GLM 4 Voice      & 3.216 & 3.38  & 2.90 & 2.787 & 2.96 & 3.80 & 3.116 & 3.27  & 2.68 & 3.040 & 3.20 & 3.13  \\
Step-Audio-AQAA  & 3.637 & 3.63  & 2.95 & 3.155 & 3.88 & 3.97 & 3.225 & 2.93  & 3.04 & 3.339 & 3.48 & 3.32  \\
Step Audio 2     & \underline{3.861} & \underline{3.93}  & 3.24 & \underline{3.189} & \underline{4.05} & 4.23 & \underline{3.479} & 3.67 & 3.32 & \underline{3.510} & \underline{3.88} & 3.60 \\
Kimi Audio       & 3.490 & 3.91  & 2.57 & 2.751 & 3.15 & 2.81 & 3.358 & \underline{3.69}  & 2.60 & 3.200 & 3.58 & 2.66  \\
GPT-4o            & \textbf{3.866} & \textbf{4.23} & 2.95 & \textbf{3.685} & \textbf{4.28} & 4.24 & \textbf{3.641} & \textbf{4.07} & 3.06 & \textbf{3.731} & \textbf{4.19} & 3.42  \\
\bottomrule
\end{tabular}
}
\label{tab:multidialogue-cqa-nvspeech}
\vspace{-0.4cm}
\end{table*}

\vspace{-0.4cm}
\subsection{Multi-Turn Interactive Evaluation Framework}
\label{subsec:multiturn}
To better evaluate the effectiveness of SDMs in real interactive conversations, we design a multi-turn interactive dialogue loop evaluation framework, as illustrated in Fig.~\ref{pic:overview}. The evaluation process begins with the construction of a user profile, which specifies the scenario, goal, and user character. To build diverse and realistic profiles, we extract user attributes from English and Chinese dialogues using GPT-4o~\cite{gpt4o} and DeepSeek-R1~\cite{guo2025deepseekr1}, respectively. We ensure topic diversity by sampling dialogues from daily-life and emotional scenarios, using LLM-based topic annotation and stratified sampling. Each instance includes a user profile and an initial dialogue utterance. The first utterance is then transformed into an emotional audio signal using Step-Audio-TTS\footnote{https://huggingface.co/stepfun-ai/Step-Audio-TTS-3B}, which serves as the initial input to the SDMs and initiates the dialogue loop.

During multi-turn interaction, the user’s text responses are generated by a chat LLM and subsequently converted into speech with emotional prompts via the TTS module. The SDM receives these audio inputs and generates both spoken and textual outputs, enabling an end-to-end audio-based conversational exchange that closely mirrors human–machine interaction. The process iterates until a termination condition is reached, such as explicit user termination, sufficient emotional relief, or repeated stagnation. To simulate the user, we adopt DeepSeek-V3.1 as the chat LLM and Step-Audio-TTS as the speech synthesizer. An additional LLM is employed to decide when the conversation should terminate, providing a dynamic and flexible evaluation loop. To improve contextual appropriateness and emotional expressiveness, we design an emotion conditioning mechanism. Given a user’s output sentence, another LLM~\cite{gpt4o} determines the most suitable emotion for the context. According to this decision, we retrieve a matching audio prompt from a curated emotional speech dataset. Specifically, we recorded 38 emotional prompts spanning diverse categories, such as sadness, fear, happiness, relaxation, excitement, humor, hesitation, and empathy. The TTS model then conditions on these prompts to generate human-like emotional speech.

For evaluation, we assess the models from both acoustic and textual perspectives to measure their ability in emotion awareness and application. Building on prior works~\cite{manku2025emergenttts} validating Audio-aware Large Language Model (ALLM)-based assessment, we adopt Gemini 2.5 Pro to score the emotional alignment of speech outputs. We design prompt engineering strategies to ensure reliable judgments, such as requiring timestamp-based textual and acoustic analyses of dialogue history, the latest user audio, and the SDM’s response, as well as incorporating final sanity-check steps before scoring. Besides, we also incorporate a text-based assessment using DeepSeek-R1 to evaluate dialogue-level EI. 
\vspace{-0.4cm}
\subsection{Data Construction}
\label{subsec:data connstruction}
\textbf{Emotion Understanding and Recognition.} 
We curated data from open source datasets UnderEmotion~\cite{yan2025uro} and NVSpeech~\cite{liao2025nvspeech}, applying an LLM-based filtering process to remove inappropriate or low-quality samples. UnderEmotion~\cite{yan2025uro}, part of the URO benchmark, contains a total of 216 dialogues, including both Chinese and English data, while NVSpeech~\cite{liao2025nvspeech} provides word-level annotations of 18 paralinguistic vocalization categories. For benchmark construction, we used edge-tts~\footnote{https://github.com/rany2/edge-tts} to generate question prompts tailored to these datasets. Examples of the constructed QA pairs are shown in Fig~\ref{pic:data_example}. 

\textbf{Emotion Reasoning and Application.}
For emotional reasoning tasks, we additionally incorporate data from StyleTalk~\cite{styletalk} and NVSpeech~\cite{liao2025nvspeech}. StyleTalk~\cite{styletalk} is a dataset where two utterances share identical content but differ in speaking style, resulting in distinct responses. Each sample contains paired data consisting of dialogue history $H$, current text and audio $(C_t, C_a)$, and the corresponding response $(R_t, R_a)$. In cases where the dialogue history and the current text remain the same, but the speaking style differs, the responses also differ, forming pairs such as $(H, C_t, C_a, R_t, R_a)$ and $(H, C_t, \hat{C_a}, \hat{R_t}, \hat{R_a})$. We construct QA tasks by asking the model to identify which response is correct, thereby testing its EI in conversational contexts. For NVSpeech~\cite{liao2025nvspeech}, we observe that even the same label, such as ah, may carry different meanings depending on context, e.g., \texttt{Question-ah} versus \texttt{Surprise-ah}. To capture this nuance, we design semi-open QA tasks that require the model to infer the intended emotion or intention behind such paralinguistic expressions. The data for the interactive dialogue task are drawn from everyday chit-chat and emotional counseling corpora, including MultiDialogue~\cite{multidialogue}, NVSpeech~\cite{liao2025nvspeech}, PsyQA~\cite{sun2021psyqa}, and PsyDTCorpus~\cite{xie2024psydt}, with detailed task construction described in Section~\ref{subsec:multiturn}. Illustrative examples of these tasks are provided in Fig.~\ref{pic:data_example}.

\vspace{-0.4cm}
\section{Experiments}
\label{sec:experiments}
\subsection{Experiment Setup}
We employ six SDMs to perform the tasks in Multi-Bench, and adopt two LLMs for evaluation: Gemini-2.5-Pro, which focuses on the acoustic dimension, and DeepSeek, which evaluates from the textual perspective. These judges are used to assess EI in multi-turn interactive dialogues. The six SDMs include GPT-4o~\cite{gpt4o}, Qwen 2.5 Omni~\cite{xu2025qwen25omnitechnicalreport}, GLM 4 Voice~\cite{zeng2024glm}, Step-Audio-AQAA~\cite{huang2025stepaudioaqaa}, Step Audio 2~\cite{stepaudio2}, and Kimi Audio~\cite{ding2025kimi}. To avoid unbounded interactions, we limit the dialogue to a maximum of ten turns.

\vspace{-0.4cm}
\subsection{Results and Analysis}
As shown in Table~\ref{tab:understanding}, GPT-4o achieves the best overall performance on tasks related to understanding and reasoning, except for emotion recognition. Step Audio 2 leads in emotion recognition with 70.80\%. Furthermore, it performs well on reasoning tasks. For example, it reaches 56.14\% accuracy in best response style inference compared with 55.29\% for Kimi Audio, and 40.00\% in paralinguistic emotion inference compared with 35.28\% for Kimi Audio. Despite these differences, multi-choice questions remain difficult for all systems, with consistently low accuracy and models typically identifying only one correct option.

In the Interactive Dialogue evaluation, we conduct multi-turn dialogue tasks on four datasets comprising a total of 1,500 dialogues. This setting yields 157,262 dialogue turns, with an average length of 8 turns per dialogue. The results are summarized in the Gemini column of Table~\ref{tab:multidialogue-cqa-nvspeech}.
In general, on Chinese data sets, the performance gap between GPT-4o and other models is relatively small. In particular, In emotion-related dialogue tasks, Step Audio2 performs on par with GPT-4o, achieving scores of 3.861 and 3.866 respectively. However, on the English datasets, GPT-4o far outperforms others, showing that current SDMs still struggle with multi-turn English dialogues. In daily topic dialogues, as shown in the NVSpeech column of Table~\ref{tab:multidialogue-cqa-nvspeech}, most models perform similarly, with the exception of GLM 4 Voice, which falls noticeably behind. In contrast, in emotion-related dialogues, as reported in the Psy column, the performance gap becomes more pronounced: GPT-4o and Step Audio2 achieve the best results, followed by Step-Audio-AQAA and Kimi Audio, while Qwen 2.5 Omni ranks lower. These results indicate that EI presents greater challenges for SDMs than casual daily conversations. Kimi Audio shows weak performance in English dialogues, often mixing Chinese and English. GLM 4 Voice scores lowest overall due to its lack of dialogue history support, underscoring the importance of conversational memory in multi-turn tasks. To validate ALLMs as judges, we compared their assessments with those of ten human evaluators under the same pipeline and instructions. The correlation of the manually scored SDMs ranking is 0.885, indicating a strong alignment between the human assessment and the model's performance.

\vspace{-0.4cm}
\section{Conclusion}
\vspace{-0.2cm}
\label{sec:conclusion}

In this work, we introduce Multi-Bench, the first benchmark for evaluating the EI of Spoken Dialogue Models in genuinely interactive, multi-turn conversations. Through a hierarchical evaluation structure and five carefully designed tasks, Multi-Bench offers a comprehensive and reproducible framework for assessing both fundamental and advanced emotional competencies. Experimental results reveal that GPT-4o demonstrates the best overall performance, followed by Step Audio 2. We observed that performance gaps between SDMs are minor in daily conversations but become more pronounced in emotion-centric dialogues, highlighting the ongoing challenges for emotional intelligence in conversational AI. We hope that Multi-Bench will serve as a rigorous resource to drive future research.

\vfill\pagebreak
\vfill\pagebreak
\bibliographystyle{IEEEbib}
\bibliography{strings,refs}

\end{document}